\documentclass[12pt,a4paper]{article}

\usepackage{amsmath}
\usepackage{amssymb}
\usepackage{amsfonts}
\usepackage{graphicx}
\usepackage{subfig}
\usepackage{caption}
\usepackage{dcolumn}
\usepackage{bm}
\usepackage[
colorlinks,linkcolor=blue,citecolor=magenta]{hyperref}
\usepackage[top=3cm,right=2cm,bottom=2.5cm,left=2cm]{geometry}
\begin{document}
\title{Massive gravity, canonical structure \\ and gauge symmetry}

\author{Zahra Molaee$^{a}$\thanks{zahra.molaei@ph.iut.ac.ir}, Ahmad Shirzad$^{a,b}$\thanks{shirzad@ipm.ir}\\
	{\it $^{a}$Department of Physics, Isfahan University of Technology, Iran}\\
	{\it$^{b}$School of Particles and Accelerators,}\\
	{\it Institute for Research in Fundamental Sciences (IPM),}\\
	{\it P.O.Box 19395-5531, Tehran, Iran}}

\maketitle
\begin{abstract}
	 Performing Hamiltonian analysis of the massive gravity \cite{mmg} in full phase space, we see that the theory is ghost free. We also see in a more clear way that this result is intrinsic of the interaction term and does not depend on the variables involved. Since no first class constraint emerges, the theory seems to lack gauge symmetry. We show that this is due to the presence of an auxiliary field, and the symmetry may be manifest in the  St\"{u}ckelberg formulation. We give the generating functional of gauge transformation in this model. 
\end{abstract}

\section{Introduction}
Massive gravity is a theory which  modifies general relativity by considering additional terms in the Einstein-Hilbert action. This theory leads ultimately to propagating a massive spin 2 particle. There is a long history behind the massive gravity, started by Pauli and Fierz in 1939 \cite{wit1}. They first presented a linear action with respect to a spin-2 field on a flat space-time background. In 1972 , Van Dam, Veltman and Zakharov  \cite{wit2,tmg}  understood that the Fierz-Pauli action in the limit of zero mass is different from the linearized general relativity.  
Then Vainshtein observed that the problem of vDVZ discontinuity may be solved by considering the full non-linear Einstein-Hilbert action accompanied by a mass term \cite{ach}. However,  Boulware and Deser showed later that a ghost appears in the theory as the sixth  degree of freedom, known as Boulware-Deser ghost \cite{park}. 

After so many efforts, in 2010 an improvement  was made by de Rham, Gabadadze and Tolley  known as (dRGT) model. They showed that the Boulware-Deser ghost can be  avoided in the decoupling limit, in a non-linear  massive gravity with  tuned coefficients \cite{nmg}. Then, it was shown that at low orders the theory is free of the Boulware-Deser ghost\cite{drgt1, drgt2}. 
   In dRGT model the reference metric is flat. 
   In 2011  F. Hassan and R. Rosen 
   presented a ghost free non-linear massive gravity action in which the flat metric was replaced by an external arbitrary metric (see Eq. (\ref{HRa}) below for their famous action)\cite{mmg}. They proved the absence of the Boulware-Deser ghost, to all orders. Furthermore, they did not need to restrict themselves to the decoupling limit \cite{HR2, mmg2,hassan}. Using the ADM formalism, they showed in the canonical structure of the model  two more second class constraints 
   	emerge which kill the Boulware-Deser ghost\cite{berg}. Their analysis was done basically in a 12-dimensional reduced phase space in which the lapse and shift functions were absent. 
   	
   	Then Kluson tried to analyze the Hamiltonian  structure of the model in full 20-dimensional  phase space \cite{Klu2}.  He found all together one first class and 7 second class constraints, leading to an 11 dimensional phase space and remaining the ghost. In fact, two more constraints, which are responsible to omit the ghost, are missing in his analysis. Hence, there is not a satisfactory canonical analysis of the model in the full 20 dimensional phase space, though there is some investigations in \cite{comeli1} in a different language.
   
    In Ref. \cite{Klu1}, adding four St\"{u}ckelberg fields to the theory, 8 first class and 2 second class constraints are  derived which agree with the desired 10-dimensional phase space of a massive spin-2 graviton. In all of the canonical investigations an intelligent change of variables made by Hassan and Rosen  seems to be essential. The most important consequence of this change of variables is that the action would be linear with respect to the lapse function. 
   	
   In this paper 
   		we first give a brief survey on the constraint structure of the model in the full 20 dimensional phase space. Some details of this analysis are reviewed in appendix A, and the main constraint structure is given in section 2. We see that the model contains 10 second class constraints. This leads to  a 10-dimensional reduced phase space of a massive spin-2 theory in 4 dimensions.
    
     We also emphasize in subsection 2.1 that the above mentioned change of variables is not the essential point in the canonical structure of the theory. In other words, the essential behavior of the theory, including the number of degrees of freedom does not depend on the variables invoked.
      
 Note, however, the main goal of doing a Hamiltonian analysis for a complicated model is not just finding the number of degrees of freedom. One also expects to study the symmetry properties of the model by considering the first class constraints of the model which act as the generators of gauge transformations in the phase space. This important task is not discussed in the literature yet. In the original model of massive gravity, although the system is symmetric under diffeomorphism, this symmetry is not showed up in the constraint structure of the model. The reason is the presence of an external field in the theory. However, in the St\"{u}ckelberg formalism the field $f_{\mu\nu}$ is no longer external. The canonical structure of massive gravity in the St\"{u}ckelberg formalism is studied in details in Refs. \cite{Klu1,HR4}. However, the role of the derived first class constraints as the gauge generators of the gauge symmetry is not clarified yet. In section 3, using valuable results of Ref. \cite{Pons} about the generating functional of  diffeomorphism, we show how the diffeomorphism gauge symmetry is generated by the first class constraints of the system. 
   
   \section{  Hamiltonian structure in full set of variables}
This section is devoted to the Hamiltonian analysis of  massive gravity in ADM formalism and with full set of canonical variables. The action is introduced in Ref. \textbf{\cite{mmg}} as 
\begin{equation}
S_{MG}=M^{2}_{P} \int d^{4}x \sqrt{-g} 
\left(R(g)+2m^{2} \sum_{n=0}^{4} \beta_{n}e_{n}(\Bbbk)\right),\label{HRa}
\end{equation}
where $\beta_{n}$ are free dimensionless parameters, $m$ is a mass parameter, $M_{P}$ is the Plank mass, the matrix  $\Bbbk \equiv \sqrt{g^{-1}f}$ is the square root of the matrix  $g^{-1}f\equiv g^{\mu\lambda}f_{\lambda\nu}$ and the elementary symmetric polynomials $e_{n}(\Bbbk)$ are as follows 
\begin{eqnarray}&&
e_{0}(\Bbbk)=1,\label{a2}
\nonumber\\ &&
e_{1}(\Bbbk)=[\Bbbk],\label{a3}
\nonumber\\ &&
e_{2}(\Bbbk)=\frac{1}{2}([\Bbbk]^{2}-[\Bbbk^{2}]),\label{a4}
\nonumber\\ &&
e_{3}(\Bbbk)=\frac{1}{6}([\Bbbk]^{3}-3[\Bbbk][\Bbbk^{2}]+2[\Bbbk^{3}]),\label{a5}
\nonumber\\ &&
e_{4}(\Bbbk)=\frac{1}{24} ( [\Bbbk]^{4}-6[\Bbbk]^{2}[\Bbbk^{2}] + 
3[\Bbbk^{2}]^{2}+8[\Bbbk][\Bbbk^{3}]-6[\Bbbk^{4}]),\label{a6}
\nonumber\\ &&
e_{i}(\Bbbk)=0,\ \ i > 4,\label{a7}
\end{eqnarray}
where  $[\Bbbk] \equiv Tr(\sqrt{g^{-1}f})$.
In ADM approach, the metric has the following (3+1) decomposition\cite{pad}
\begin{eqnarray} &&
g_{00}=-N^{2}+N_{i}N^{i},\hspace{7mm} g_{0i}=N_{i},\hspace{7mm} g_{ij}=\gamma_{ij},\label{a8} \nonumber\\ &&
g^{00}=\frac{-1}{N^{2}},\hspace{7mm} g^{0i}=\frac{N^{i}}{N^{2}},\hspace{7mm}g^{ij}=\gamma^{ij}-\frac{N^{i}N^{j}}{N^{2}},\label{a9}
\end{eqnarray}
where $N, N^{i}$ are  lapse and shifts functions respectively, so the indices are raised and lowered by the spatial metric $ \gamma_{ij} $.
The background metric  $f_{\mu\nu} $ , as an auxiliary field, may be similarly written in the corresponding variables as  
\begin{eqnarray} &&
f_{00}=-M^{2}+M_{i}M^{i},\hspace{7mm}f_{0i}=M_{i},\hspace{7mm}f_{ij}=f_{ij},\label{a10} \nonumber\\ &&
f^{00}=\frac{-1}{M^{2}},\hspace{7mm} f^{0i}=\frac{M^{i}}{M^{2}},\hspace{7mm}f^{ij}=f^{ij}-\frac{M^{i}M^{j}}{M^{2}}.\label{a11}
\end{eqnarray}
The possibility of finding coordinate systems which lead to simultaneous ADM decomposition of the metrics $ g_{\mu\nu} $ and $ f_{\mu\nu} $ is not an obvious fact. However, in Ref. \cite{HK} it is showed that in fact it is possible to have a common ADM splitting for $ g $ and $ f $. 
The Lagrangian density associated with the action (\ref{HRa}) reads
\begin{equation}
\mathcal{L}=\pi^{ij} \partial_{t} \gamma_{ij}+N R_{0}+N^{i} R_{i}+2m^{2}N\sqrt{\gamma} \sum_{n=0}^{4} \beta_{n} e_{n}(\sqrt{g^{-1}f}),\label{a12}
\end{equation}
where $R_{0}$ and $R_{i}$ are as follows\cite{pad}
\begin{equation}
R_{0}=\sqrt{\gamma} \ ^{3}R+\dfrac{1}{\sqrt{\gamma}}\left(\frac{1}{2}\pi^{2}-\pi^{ij}\pi_{ij}\right),\hspace{10mm} R_{i}=2\sqrt{\gamma}\gamma_{ij} \triangledown _{k}\left(\frac{\pi^{jk}}{\sqrt{\gamma}}\right).\label{a14}
\end{equation}
As is apparent, the time derivative of lapse and shift functions is absent from the Lagrangian (\ref{a12}). This enables us to consider $N$ and $N^{i}$ as auxiliary fields, and use just $\gamma_{ij}$ and the corresponding momenta $\pi^{ij}$ as the canonical variables, where 
\begin{equation}
\pi^{ij}=-\sqrt{\gamma}(K^{ij}-g^{ij}K)\label{c1},
\end{equation}
 in which $K_{ij}$ is the extrinsic curvature of the surface $ t $$ = $constant. The authors of Ref. \cite{mmg2} used this approach and worked in a 12-variables phase space. Although it may seem simpler to work in a smaller phase space, however, one should be careful about the dependence of $N$ and $N^{i}$ on canonical variables in calculating the Poisson brackets. To avoid such miscalculations we prefer to consider the full set of variables  $N$, $N^{i}$, $\gamma_{ij}$ and the corresponding momenta as a set of 20 canonical variables. 
It can be shown that the action \ref{a12} in the set of Hassan and Rosen variables is as follows
 \begin{equation}
 \mathcal{L}=\pi^{ij} \partial_{t} \gamma_{ij}+ M n^{i}R_{i}+M^{i}R_{i}+2m^{2}\sqrt{\gamma}M K+N(R_{0}+R_{i}D_{j}^{i}n^{j}+2m^{2}\sqrt{\gamma}Z),\label{HRL2}
 \end{equation}
 where the functions $ K $ and $ Z $ are introduced in appendix A, Eqs. (\ref{INT1}) and (\ref{INT2}). 
 The primary constraints are  $ P $ and $p_{i}$ associated to $ N $ and $n_{i}$. Time
 evaluation of the system gives the secondary constraints as $\chi$ , $\vartheta_{i}$ given in Eqs. (\ref{dtp1}) and (\ref{vk}). The constraints  $\vartheta_{i}$ are the same equations derived in \cite{mmg2}  as the equations of motion $n_{i}$. On the constraint surface $\vartheta_{i}=0$, $ n^{i} $'s are determined in terms of $ \pi^{ij} $
and $ \gamma_{ij} $. 
 Since $\vartheta_{i}$ depend on $ n^{i} $, the Poisson brackets $ \{\vartheta_{i}, p^{i} \} $ are non-zero. So consistency condition of $ \vartheta_{i} $'s leads to determining the Lagrange multipliers $u_{i}$, associated to the primary constraints $ p_{i} $. Hence, the six second class constraints $p_{i}$ and $ \vartheta_{i} $ are put aside from our investigation.

 Consistency of $ \chi $ gives a new constraint $ \varUpsilon $ given in Eq. (\ref{op}). The important point is that the new constraint $ \varUpsilon $ is independent of $N$. Thus, consistency condition of  $ \varUpsilon $ gives another  constraint
 $ \varOmega\equiv \lbrace \varUpsilon, H_{0}\rbrace_{*}+N \lbrace \varUpsilon,\chi \rbrace_{*}  \approx 0 $. Since $ \varOmega $
 depends linearly on $ N $, consistency of $ \varOmega $
 determines ultimately the remaining Lagrange multiplier
$  u $ in the total Hamiltonian $H_{T}$ of Eq. (\ref{c2}).

In this way we derived a four level chain of second class constraints $(P_{N}\rightarrow \chi\rightarrow\varUpsilon\rightarrow\varOmega) $. Adding them to the three two-level second class constraints $(P_{i}\rightarrow \vartheta^{i})$, the total number of constraints is 10. Subtracting from the total number of degrees of freedom\cite{zms2} we have $20-10=10$ dynamical fields in the model. Divided by 2, we have 5 dynamical fields in the Lagrangian formulation, corresponding to a massive spin-2 theory. So, there is no room for appearing further degrees of freedom to be considered as Boulware-Deser ghost.

As we see, the most advantage of the massive gravity model of Hassan-Rosen is exhibited by appearing two more constraints $\varOmega$ and $\varUpsilon$ which are responsible for omitting the extra degree of freedom, i.e. the Boulware-Deser ghost. This has happened due to linearity of Lagrangian (\ref{HRL}) with respect to lapse function $N$ after  changing of variables from $N_{i}$ to $n_{i}$. However, in what follows we show that the change of variables from $N_{i}$ to $n_{i}$ is not the essential point of the dynamical analysis.

\subsection{ Independence from the choice of variables}
 The main point about the Lagrangian given by Hassan-Rosen is that the given interacting term (i.e. $ V=\sum \beta_{n}e_{n}(\sqrt{g^{-1}f})$), when written in terms of $N$ and $n_{i}$ is linear with respect to $N$. We can assert that the matrix of second partial derivatives of $V(N,n_{i},\gamma_{ij})$ with respect to the variables $N$ and $n_{i}$ is singular. In fact it has rank 3, instead of being full rank. 
 As pointed out in Refs. \cite{drgt1}, \cite{HR2}, \cite{mmg2} and \cite{comeli1} this characteristics is the main property of the well-tuned interaction term in Eq. (\ref{HRa}) which leads the theory to being ghost free.
 To see the point, consider the most general form of the Hilbert-Einstein Lagrangian plus an interaction term which does not depend on the derivatives of the metric. Defining four variables $N^{\mu}$ as ($N,N^{i}$), the corresponding Lagrangian can be written similar to Eq. (\ref{a12}) as 
\begin{equation}
\mathcal{L}=\dot{\gamma_{ij}}\pi^{ij}-H_{c}(N^{\mu},\gamma_{ij},\pi^{ij}),
\end{equation}
where the canonical Hamiltonian for a general interaction is
\begin{equation}
H_{c}=NR_{0}(\gamma,\pi)+N^{i}R_{i}(\gamma,\pi)+V(N^{\mu},\gamma).
\end{equation}
The momenta $P_{N^{\mu}}$ appear as four primary constraints and
the total Hamiltonian reads,
\begin{equation}
H_{T}=H_{c}+u_{\mu}P_{N^{\mu}}, \label{total}
\end{equation}
The consistency condition on the time evolution for $P_{N^{\mu}}$ gives
\begin{equation}
\zeta^{\mu}=\lbrace P_{N^{\mu}},H_{T} \rbrace=\frac{\partial H_{c}}{\partial N^{\mu}}\approx 0. \label{zeta1}
\end{equation}
Consistency of the secondary constraints reads as $ \lbrace \zeta_{\mu},H_{T} \rbrace \approx 0$. This leads to the following equations for the Lagrange multipliers $u^{\mu}$:
\begin{equation}
  \lbrace \zeta_{\mu},P_{N^{\nu}} \rbrace u^{\nu}\approx-\lbrace \zeta_{\mu},H_{c} \rbrace . \label{zeta2} 
\end{equation}
Using Eq. (\ref{zeta1}) we can write Eq. (\ref{zeta2}) as follows 
\begin{equation}
\frac{\partial^{2} V}{\partial N^{\mu} \partial N^{\nu}} u^{\nu}=-\lbrace \frac{\partial H_{c}}{\partial N^{\mu}} ,H_{c} \rbrace. \label{Consis}
\end{equation}
If the interaction potential $V$ is chosen so that $\partial^{2} V/\partial N^{\mu} \partial N^{\nu}$ is non-singular, then all Lagrange multipliers $u^{\nu}$ would be determined. This means that the set of 8 constraints $P_{N^{\mu}}$ and $\zeta_{\mu}$ are second class. 
The reduced phase space of the system can be achieved by imposing strongly the constraints $ P_{N^{\nu}}=0 $ and $ \zeta_{\mu}=0 $. The latter equations determine the functions $ N^{\mu} $ in terms of the canonical variables $\gamma_{ij},\pi_{ij}  $. This is in agreement of treating the problem in a 12 dimensional phase space and considering the lapse and shift functions as auxiliary fields.

So the number of dynamical variables is $20-8=12$,
in phase space which means 6 degrees of freedom in configuration space. This can be interpreted as existing a massive spin-2 graviton in plus a scalar ghost. 

Hence, to avoid the Boulware-Deser ghost we need a potential term for which the matrix of second order derivatives with respect to $N^{\mu}$  is  singular. Suppose $\partial^{2} V/\partial N^{\mu} \partial N^{\nu}$ is a $ 4\times 4$  matrix with ranks 3. We want to show that this property remains valid if we change the variables $N^{\mu}$ to $n^{\mu}$, say $(N,n^{i})$. It is easy to see that 
\begin{equation}
 \dfrac{\partial^{2}H_{c}}{\partial n^{\mu}\partial n^{\nu}} = 
 \dfrac{\partial^{2}H_{c}}{\partial N^{\alpha}\partial N^{\beta}} \dfrac{\partial N^{\alpha}}{\partial n^{\mu}}\dfrac{\partial N^{\beta}}{\partial n^{\nu}} +  \dfrac{\partial H_{c}}{\partial N^{\alpha}}  \dfrac{\partial^{2}N^{\alpha}}{\partial n^{\mu}\partial n^{\nu}} . \label{HC1}
\end{equation}
The second term in the right hand side of Eq. (\ref{HC1}) vanishes due to the constraints $\zeta_{\mu}$ in Eq. (\ref{zeta1}). Assuming the Jacobian $\partial N^{\alpha}/\partial n^{\mu}$ to be nonsingular, we deduce that the rank of 
$\partial^{2}H_{c}/\partial n^{\mu}\partial n^{\nu}$ is the same as $\partial^{2}H_{c}/\partial N^{\alpha}\partial N^{\beta}$.

In the current model of action (\ref{HRa}) the ability of the interaction $ \sqrt{g^{-1}f} $ to be linearized with respect to N makes us sure that the matrix $\partial^{2} V/\partial N^{\mu} \partial N^{\nu}$  has rank 3 in every set of variables $(N,N^{i},\gamma_{ij})$. However, it can be shown vice versa that if $rank \mid\partial^{2} V/\partial N^{\mu} \partial N^{\nu}\mid=3$ one can in principle find a set of variables  $(N,N^{i},\gamma_{ij})$ such that the interaction term is linear with respect to $N$.

Hence, if one does not change the variables, nevertheless singularity of $\partial^{2}H_{c}/\partial N^{\alpha}\partial N^{\beta}$ plays its role in the Hamiltonian structure of the system. Noticing to consistency equations (\ref{Consis}) for the second level constraints $\zeta_{\mu}$, we see that one combination of the Lagrange multiplier $u^{\nu}$ say $\tilde{u}$ would remain undetermined. This means that there exist one combination of $\zeta_{\mu}$ which has been remained as a first class constraint so far, 
i.e. up to this level of consistency; although it may change to second class in the subsequent steps, as we will see. Denote this constraint as $\tilde{\chi}(N^{\mu},\gamma_{ij},\pi^{ij})$. 

For $\tilde{\chi}$, the left hand side of Eq. (\ref{Consis}) vanishes and the right hand side may be considered as the third level constraint $\tilde{\varUpsilon}(N^{\mu},\gamma_{ij},\pi^{ij})$. 
Let us emphasis at this point that, in principal, the matrix  $\partial^{2} V/\partial N^{\mu} \partial N^{\nu}  $  as well as its null-vector depend on the variables $ N^{\mu} $, and this is the case for the  third level constraint $ \tilde{\varUpsilon} $ . Hence,  $ \{\tilde{\varUpsilon} ,P_{N^{\mu}}\} $ does not vanish and may lead to determining the remaining undetermined Lagrange multiplier $\tilde{u}$, by going through consistency of the constraint $\tilde{\varUpsilon}$ using the total Hamiltonian (\ref{total}). This point is also discussed more or less in \cite{comeli1}. In this way we have found all together 9 second class constraints $ P_{N^{\mu}} $ , $ \zeta_{\mu} $ and $ \tilde{\varUpsilon}$ which gives a phase space with 11 dynamical fields, as stated in \cite{Henu},  this should not be understood as an odd dimensional phase space, since in fact we have infinite number of degrees of freedom in a field theory. 

Suppose  $  \tilde{\varUpsilon}$   depend on $ N^{\mu} $ in such a way that it does not lead to determination of the remaining Lagrange multiplier in the consistency procedure. For instance, for the case where the interaction is linear with respect to $ N $, as we discussed in previous section, the constraint $ \tilde{\varUpsilon} $ does not depend on  $ N  $ due to  $\tau^{i}\approx 0$ (see Eq. (\ref{equal2})). Then, consistency of $\tilde{\varUpsilon}$ gives, in principle, another constraint $\tilde{\varOmega}$, which gives rise to determining the Lagrange multiplier $\tilde{u}$ upon consistency condition. 
In this way we have 10 second class constraints and no arbitrary function in the theory. It worth noting that the variables $ N^{\mu} $ are determined, on the constraint surface, in terms of the canonical variables. In fact, the constraints  $\zeta_{\mu}$  determines three of them, since the rank of $ \partial \zeta_{\mu} /\partial N^{\mu}=\partial^{2} V/\partial N^{\mu} \partial N^{\nu}  $  is three, and the forth one is determined through strongly imposing the constraint $\tilde{\varOmega}=0$.  For example in the framework used in the previous section,  the variables $n^{i}$ are determined through imposing the constraints $ \vartheta_{i}=0 $ (Eq. (\ref{vk}))
and the lapse function $ N$ is determined as the result of  $\varOmega$ in Eq. (\ref{omega}). 
Unfortunately the corresponding expressions of the above mentioned constraints
for the Lagrangian (\ref{a12}) in the original variables are very complicated due to square-root term $\sqrt{g^{-1}f}$ in the interaction term. However, our observation of uniqueness of the rank of $\partial^{2}H_{c}/\partial N^{\alpha}\partial N^{\beta}$ shows that the result is the same as what we investigated in the new variables. We conclude that the essential point is not the choice of dynamical variables. Instead, the main point is the singular behavior of the matrix $\partial^{2}V/\partial N^{\alpha}\partial N^{\beta}$.
 
\section{ Diffeomorphism and St\"{u}ckelberg formalism}
As we observed, in the Hamiltonian structure of massive gravity resulted to Eq. (\ref{HRL}), all the constraints are second class. On the other hand, we know from Dirac theory of constrained systems that in every gauge theory the gauge transformations are generated by first class constraints. Now the question arises how the Hamiltonian constraint structure of the system may support the diffeomorphism symmetry contained in the model. In other words, we expect a system of first class constraints as the generators of diffeomorphism gauge symmetry. The point is that in the Lagrangian of the theory we have inserted the external fields $f_{\mu\nu}$ as a second rank tensor with the pre-determined dynamics. Hence, a diffeomorphism transformation on the dynamical variables $N, N_{i}$ and $g_{ij}$ should be accompanied by  definite variations of external fields $f_{\mu\nu}$. Therefore, the dynamical content of the theory, as it stands, can not introduce a set of variations which maintain the action invariant; i.e. the invariance of the action is only guaranteed by introducing  variations of $f_{\mu\nu}$  by hand. To clarify the point, let us consider the simple example of a complex scalar field coupled to the external gauge field $A_{\mu}$ as

\begin{equation}
\mathcal{L}=((\partial^{\mu}-ieA^{\mu})\phi^{+} (\partial_{\mu}+ieA_{\mu})\phi-m^{2}\phi^{+}\phi)=D_{\mu}\phi D^{\mu}\phi^{+}-m^{2}\phi^{+}\phi. \label{Amunu}
\end{equation}
This model is symmetric under the gauge transformation $\phi\longrightarrow e^{i\varLambda(x)}\phi$ provided that the external field undergo simultaneously the gauge transformation $A_{\mu}\longrightarrow A_{\mu}-\frac{1}{e}\partial_{\mu}\varLambda$. The canonical structure of the system implies two canonical momentum fields as
\begin{equation}
\Pi^{+}=\dot{\phi}+ieA^{0}\phi,\hspace{15mm}\Pi=\dot{\phi}^{+}-ieA_{0}\phi^{+},
\end{equation}
which do not introduce any constraint. Hence, there is no first class constraint to generate gauge transformation, although we know that the theory is gauge symmetric. This is because of the existence of external fields with gauge transformations imposed by hand. However, if we add dynamical kinetic term for the gauge field (i.e. $-\frac{1}{4}F_{\mu\nu}F^{\mu\nu}$), then we would have a gauge system in which the gauge transformations of the scalar fields $\phi$ and $\phi^{+}$ as well as the gauge field $A_{\mu}$ would be resulted automatically from the first class constraints of the system.

The massive gravity is also formulated in St\"{u}ckelberg formalism, in which four scalar fields $\phi^{A}$ are considered as St\"{u}ckelberg fields. Let us restrict ourselves to  $\beta_{1}$-model where $\beta_{1}=1$ and the remaining $\beta_{n}$' s in action (\ref{HRa}) vanish. The action is given by 

\begin{equation}
S=M^{2}_{P}\int d^{4}x \sqrt{-g}[R-m^{2}Tr\sqrt{A}], \label{stmg}
\end{equation}
 where 
 \begin{equation}
 A^{A}_{B}\equiv \partial_{\mu}\phi^{A}g^{\mu\nu}\partial_{\nu}\phi^{C}\bar{f}_{CB},
 \end{equation}
in which $A,B,\cdots$ are Lorentzian indices and $\bar{f}_{CB}$ is a background field such as $\eta_{CB}=diag(-1,1,1,1)$. 
It is shown \cite{HR4} that this formulation is equivalent to the original model of the action (\ref{HRa}). The canonical structure of massive gravity in the St\"{u}ckelberg formulation is discussed in Refs. \cite{HR4} and \cite{Klu1}. Here, we review very briefly the latter work, in order to use the results in what follows. Using the ADM variables, the action (\ref{stmg}) can be written as, 
 \begin{equation}
 S=M^{2}_{P}\int d^{4}x \left( \dot{\gamma}_{ij}\pi_{ij}+\dot{\phi}^{A}P_{A}-N(H_{0}^{GR}+H_{0}^{ST})-N^{i}(H_{i}^{GR}+H_{i}^{ST})\right),\label{stu}
 \end{equation}
where
\begin{equation}
P_{A}=\frac{\delta L}{\delta \partial_{t}\phi^{A}}=M^{2}_{P}m^{2}\sqrt{g}(A^{-1/2})_{AB} \triangledown_{n}\phi^{B}.
\end{equation}
The additional terms $H_{0}^{ST}$ and $H_{i}^{ST}$ emerge due to interaction term $m^{2}Tr\sqrt{A}$ in the action (\ref{stu}) as follows
\begin{eqnarray}&&
H_{0}^{ST}=m^{2}M_{P}^{2}\sqrt{g} (\Pi^{AB}g^{ij} \partial_{i}\phi^{A}\partial_{j}\phi^{B})^{1/2}\nonumber\\&&
H_{i}^{ST}=P_{A}\partial_{i}\phi^{A},
\end{eqnarray}
where
\begin{equation}
\Pi_{AB}\equiv\frac{1}{gm^{4}M_{P}^{4}} P_{A}P_{B}+\eta_{AB}.
\end{equation}
As is apparent, the momenta $P_{N}$ and $P_{N^{i}}$, conjugate to $N$ and $N_{i}$ are primary constraints. Consistency of $P_{N}$ and $P_{N^{i}}$ gives the secondary constraints  $H_{0}=H_{0}^{GR}+H_{0}^{ST}$ and $H_{i}=H_{i}^{GR}+H_{i}^{ST}$. As given in full details in Ref. \cite{Klu1} all of these 8 constraints are first class. On the other hand, there exists another primary constraint as 
\begin{equation}
C=1+\frac{1}{gm^{4}M_{P}^{4}}P_{A}P^{A},
\end{equation}
which gives, under consistency, the secondary constraint $NC_{2}$ where
\begin{equation}
C_{2}=P_{A}\partial_{i}\Pi^{AB}\sqrt{g}\partial_{j}\phi_{B}(\sqrt{\tilde{\Pi}^{-1}})^{ij}-\frac{2}{\sqrt{g}m^{4}M_{P}^{4}} P_{A}P^{A}g_{ij}\pi^{ij},
\end{equation}
in which
\begin{equation}
\tilde{\Pi}=g^{ij}\partial_{j}\phi_{A}\Pi^{AB}\partial_{i}\phi_{B}.
\end{equation}
Since the lapse function can not vanish for a physical metric, the system possesses two more constraints $C$ and $C_{2}$ which are second class with respect to each other. Considering 8 first class and 2 second class constraints for a system of 28 degrees of freedom we have ultimately 10 dynamical degrees of freedom in the reduced phase space which is equivalent to a massive spin-2 system and no ghost exists. 

Comparing to the original model of massive gravity which contains no first class constraint, here we have 8 first class constraints in two generations. This makes us hopeful to write the generator of gauge transformations as a combination of first class constraints and the gauge parameters. For this reason let us recall the main results of a precise method due to Pons, Salisbury, Shepley \cite{Pons} concerning the generator of diffeomorphism in some special covariant systems. These systems are generalizations of the GR where in the ADM formalism the canonical Hamiltonian is given by $H=N_{\mu}H^{\mu}$ such that $H^{\mu}$ are four first class functions of the canonical fields. They show the following gauge generator produces all of the variations due to diffeomorphism 
\begin{equation}
G(t)=P_{\mu}\dot{\xi}^{\mu}+(H_{\mu}+N^{\rho}C^{\nu}_{\mu\rho}P_{\nu})\xi^{\mu}, \label{Gen}
\end{equation}
where the coefficients $C^{\sigma}_{\mu\nu}$ are defined in the following Poisson brackets
\begin{equation}
\left\lbrace H_{\mu},H_{\nu} \right\rbrace = C^{\sigma}_{\mu\nu}H_{\sigma},
\end{equation}
and the arbitrary infinitesimal parameters $\xi^{\mu}$ are related to infinitesimal coordinate transformations $\epsilon^{\mu}$ (i.e. $x^{\mu}\longrightarrow x^{\mu}-\epsilon^{\mu}$) as follows
\begin{equation}
\epsilon^{0}=\frac{\xi^{0}}{N},\hspace{10mm} \epsilon^{i}=\xi^{i}-\frac{N^{i}\xi^{0}}{N}. \label{epsilon}
\end{equation}
Note that a spatial integration should be understood whenever the indices contract. For our case of massive gravity in St\"{u}ckelberg formulation, we are happy to see that it fulfills the conditions of the systems discussed in \cite{Pons}. The coefficients $C^{\sigma}_{\mu\nu}$ can be seen to be the same coefficients of GR as 
\begin{eqnarray} &&
C^{i^{\prime \prime}}_{00^{\prime}}=g^{ij}(x^{\prime \prime})(\delta^{3}(x-x^{\prime\prime})+\delta^{3}(x^{\prime}-x^{\prime\prime}))\frac{\partial \delta^{3}(x-x^{\prime})}{\partial x^{j}},
\nonumber \\ &&
C^{0^{\prime \prime}}_{i0^{\prime}}=\delta^{3}(x-x^{\prime \prime})
\frac{\partial\delta^{3}(x-x^{\prime})}{\partial x^{i}}= -C^{0^{\prime \prime}}_{0^{\prime} i},
\nonumber \\ &&
C^{k^{\prime\prime}}_{i j^{\prime}}=(\delta^{k}_{i}\delta^{3}(x^{\prime\prime}-x^{\prime})\frac{\partial}{\partial x^{j}}+\delta^{k}_{j}\delta^{3}(x^{\prime\prime}-x)\frac{\partial}{\partial x^{i}})\delta^{3}(x-x^{\prime}).
\end{eqnarray}
Putting these coefficients in Eq. (\ref{Gen}) we find  
\begin{eqnarray} &&
G(t) =\int d^{3}x P_{N}\dot{\xi}^{0}+P_{N^{i}}\dot{\xi}^{i}+H_{0}\xi^{0}+H_{i}\xi^{i}\nonumber \\ && \hspace{10mm}
+\int d^{3}x P_{N^{i}}(x)\xi^{0}(x) g^{ij}(x)\frac{\partial N(x)}{\partial x^{j}}+\int d^{3}x \xi^{0}(x) \frac{\partial N(x)P_{N^{i}}(x)g^{ij}(x)}{\partial x^{j}}\nonumber \\ && \hspace{10mm}
 -\int d^{3}x P_{N}(x)\xi^{0}(x) \frac{\partial N^{i}(x)}{\partial x^{i}} +\int d^{3}x P_{N}(x)\xi^{j}(x) \frac{\partial N(x)}{\partial x^{j}}\nonumber \\ &&
\hspace{10mm} + \int d^{3}x \xi^{i}(x) \frac{\partial P_{N^{i}}(x)N^{j}(x)}{\partial x^{j}}+\int d^{3}x P_{N^{j}}(x)\xi^{i}(x)  \frac{\partial N^{j}(x)}{\partial x^{i}}
\end{eqnarray}
Using the above generating functional for the variations
 of fields (i.e. $\delta\psi=\left\lbrace \psi,G \right\rbrace$) by changing the arbitrary gauge parameters as in Eq.  (\ref{epsilon}) we have the following variations of the corresponding variables 

\begin{eqnarray}&&
\delta N= N_{,\mu}\epsilon^{\mu}+N\epsilon_{,0}^{0}-NN^{i}\epsilon^{0}_{,i}, \\ &&
\delta N^{i}= N^{i}_{,\mu}\epsilon^{\mu}+N^{i}\epsilon_{,0}^{0}-(N^{2}g^{ij}+N^{i}N^{j})\epsilon^{0}_{,j}+\epsilon^{i}_{,0}-N^{j}\epsilon^{i}_{,j}, \\ &&
\delta g_{ij}= \epsilon^{0} \dot{g_{ij}}+2g_{ik} \partial_{j}\epsilon^{k}+2N_{i} \partial_{j}\epsilon^{0}+\epsilon^{e}\partial_{e}g_{ij},
 \\ &&
\delta \phi^{A}= \epsilon^{\mu} \partial_{\mu}\phi^{A}.
\end{eqnarray}
The first three lines are consistent with the covariant variation    of the metric, and the last line gives the gauge transformations of the St\"{u}ckelberg fields. 

\section{Conclusions }
In this paper we studied  the Hamiltonian structure of massive gravity in the full phase space with 20 variables. We observed that elimination of the Boulware-Deser ghost due to two more second class constraints originates from singularity of the matrix of second derivatives of the interaction term with respect to lapse and shift functions, and does not depend on the change of variables.  linearization of the action  of massive gravity with respect to the lapse function is, in fact, a powerful tool to show this property explicitly. 

Finally we studied the diffeomorphism symmetry of the massive gravity. We showed that in the original theory this symmetry is not manifest in the constraint structure of the theory. This is because all of the constraints are second class. On the other hand,  diffeomorphism symmetry in the Lagrangian level is established by assuming appropriate variations for the external field. 

In St\"{u}ckelberg formalism, however, there is no external field and all of the variables gain their dynamics through the variation of the action of theory. We showed the explicit role of the first class constraints in this formalism in generating the diffeomorphism gauge symmetry of the model. As is known, the massive gravity, although an important model in its own right,  leaded to the famous model of Bi-Gravity afterwards \cite{ HRB}. Investigation the detailed Hamiltonian analysis of Bi-Gravity is the subject of our future work. 
\\

{\bf{Acknowledgements:}} The authors would like to thank M. Alishahiha, A. Naseh and M. Hajihashemi for helpful discussions.
\vskip .3cm
\appendix
\numberwithin{equation}{section}

\section{Review of Hamiltonian structure of Massive Gravity} \label{sec:rev}

Using the change of variables used by Hassan and Rosen to make the action (\ref{HRa}) linear with respect to the lapse function. To do this, let introduce the vector $n^{i}$ as     
\begin{equation}
N^{i}=M n^{i}+M^{i}+N D_{j}^{i}n^{j},\label{a15}
\end{equation}
where
\begin{eqnarray}
&&\sqrt{x}D^{i}_{j}=\sqrt{(\gamma^{il}-D^{i}_{k}n^{k}D^{l}_{e}n^{e})f_{lj}}\ ,\label{a16} \\ &&
x=1-n^{i}f_{ij}n^{j}. \label{a17}
\end{eqnarray}
The $3\times 3$ matrix $D^{i}_{j}$ is derived as 

\begin{equation}
D^{i}_{j}=\sqrt{\gamma^{id}f_{dm}W^{m}_{n}}(W^{-1})^{n}_{j},\label{a18}
\end{equation}
where 
\begin{equation}
W^{l}_{j}=[1-n^{k}f_{km}n^{m}]\delta^{l}_{j}+n^{l}f_{mj}n^{m}.\label{a19}
\end{equation}
Then it can be shown that the action (\ref{a12}) in the new set of variables $N, n^{i},\gamma_{ij}$ and the corresponding momenta, $P, p_{i}, \pi^{ij}$ is as follows
\begin{equation}
\mathcal{L}=\pi^{ij} \partial_{t} \gamma_{ij}+ M n^{i}R_{i}+M^{i}R_{i}+2m^{2}\sqrt{\gamma}M K+N(R_{0}+R_{i}D_{j}^{i}n^{j}+2m^{2}\sqrt{\gamma}Z),\label{HRL}
\end{equation}
where the functions $K$ and $Z$ are emerged from the interaction terms as
\begin{eqnarray} &&
K=\beta_{1}\sqrt{x}+\beta_{2}(\sqrt{x}^{2}D^{i}_{i}+n^{i}f_{ij}D_{k}^{j}n^{k})\nonumber\\ &&\hspace{10mm}+\beta_{3}(\sqrt{x}(D^{l}_{l}n^{i}f_{ij}D^{j}_{k}n^{k}-D^{i}_{k}n^{k}f_{ij}D^{j}_{l}n^{l})+
\sqrt{x}^{3}\frac{1}{2}(D^{i}_{i}D^{j}_{j}-D^{i}_{j}D^{j}_{i})), \label{INT1}\\ &&
Z=\beta_{0}+\beta_{1}\sqrt{x}D^{i}_{i}+\frac{1}{2}\beta_{2}(\sqrt{x}^{2})(D^{i}_{i}D^{j}_{j}-D^{j}_{i}D^{i}_{j}) \nonumber\\ &&\hspace{15mm}+\beta_{3}\sqrt{x}^{3} \frac{1}{6} 
(D^{i}_{i}D^{j}_{j}D^{k}_{k}-3D^{i}_{i}D^{k}_{j}D^{j}_{k}+2D^{i}_{j}D^{j}_{k}D^{k}_{i}.\label{INT2}
\end{eqnarray}
The canonical momenta $P$ and $p_i$, conjugate to $N$ and $n^{i}$ respectively, appear as primary constraints and the momenta $\pi^{ij}$ are given in Eq. (\ref{c1}) in terms of the velocities $\dot{\gamma}_{ij}$.
Time evaluation of the system is due to the total Hamiltonian $H_{T}=\int d^{3}x\mathcal{H}_{T}$ where
\begin{equation}
\mathcal{H}_{T}=\mathcal{H}_{c}+uP+u_{i}p_{i}. \label{c2}
\end{equation} 
The canonical Hamiltonian density $\mathcal{H}_{c}$ is deduced from Eq. (\ref{HRL}) as 
\begin{equation}
\mathcal{H}_{c}=\mathcal{H}_{0}+N\chi,
\end{equation}
in which
\begin{equation}
\mathcal{H}_{0}= M n^{i}R_{i}+M^{i}R_{i}+2m^{2}\sqrt{\gamma}M K,
\end{equation}
\begin{equation}
\chi \equiv -(R_{0}+R_{i}D_{j}^{i}n^{j}+2m^{2}\sqrt{\gamma}Z). 
\end{equation}
The primary constraints should be preserved during the time. Using the fundamental Poisson brackets 
\begin{eqnarray} &&
\lbrace N(x),  P_{N}(y)\rbrace=\delta(x-y), \nonumber\\ &&
\lbrace n^{i}(x), P_{j}(y)\rbrace=\delta^{i}_{j}  \delta(x-y), \nonumber\\ &&
\lbrace g_{ij}(x),\pi^{kl}(y)\rbrace=\frac{1}{2}(\delta^{k}_{i}\delta^{l}_{j}+\delta^{l}_{i}\delta^{k}_{j})\delta(x-y),\label{POb}
\end{eqnarray}
we demand
\begin{eqnarray} &&
\partial_{t} P=\lbrace P,H_{T} \rbrace  \approx 0 ,\label{dtp} \\ &&
\partial_{t} p_{i}= \lbrace p_{i},H_{T} \rbrace  \approx 0 ,\label{dtpi}
\end{eqnarray}
This leads to the following relations 
\begin{eqnarray} &&
\chi \approx 0,\label{dtp1} \\ &&
\left(M \delta^{k}_{i}+N \frac{\partial(D^{k}_{j}n^{j})}{\partial n^{i}}\right) \vartheta_{k}  \approx 0,\label{dtpi1}
\end{eqnarray}
where
\begin{eqnarray} &&
\vartheta_{k}= R_{k}-2m^{2}\sqrt{\gamma}\frac{n^{b}f_{ba}}{\sqrt{x}}[\beta_{1}\delta^{a}_{k}+\beta_{2}\sqrt{x}(\delta^{a}_{k}D^{j}_{j}-D^{a}_{k}) \nonumber \\ &&
\hspace{12mm}+\beta_{3}\sqrt{x}^{2}(\frac{1}{2}\delta^{a}_{k}(D^{j}_{j}D^{i}_{i}-D^{j}_{i} D^{i}_{j})+D^{a}_{j}D^{j}_{k}-D^{a}_{k}D^{j}_{j})].\label{vk}
\end{eqnarray}
Since the matrix  $\left(M \delta^{k}_{i}+N \frac{\partial(D^{k}_{j}n^{j})}{\partial n^{i}}\right)$ in Eq. (\ref{dtpi1}) is the Jacobian of the invertible transformation (\ref{a15}), it is non singular. Hence, Eq. (\ref{dtpi1}) gives $\vartheta_{k} \approx 0$. Since $\vartheta_{k}$ are three independent functions of $n_i$ the set of 6 constraints $p_{i},\vartheta_{k} $ constitute a second class system, which according to the Dirac approach, can be imposed after all as strong equalities\cite{dms3}. Hence, the term $u_{i}p^{i}$ in the total Hamiltonian should be omitted. Also by using $\vartheta_{k} = 0$ we can determine $ n^{i} $ in terms of $\pi^{ij}$ and $ \gamma_{ij}$. This point implies some subtleties in the following calculations which should be taken into account carefully.

In fact, whenever we encounter second class constraints, we should change the Poisson brackets to Dirac brackets afterwards. To do this, it is simpler to consider the variables $n^{i}$ as functions of $\gamma^{ij}$ and $\pi^{ij}$. So, from now on, every Poisson bracket should also include the dependence of $n^{i}$ on the canonical variables.

Eq. (\ref{dtp1}) implies $\chi$ as the second level constraint follows  the primary constraint $P$.
Using the total Hamiltonian density $\mathcal{H}_{T}=\mathcal{H}_{0}+N\chi+uP$, the
Consistency of  $\chi(x)$  gives\footnote{If a function $\psi$  depends explicitly on time, then its time derivative is obtained via  $\dot{\psi}=\{\psi, H_{T}\} +\frac{\partial \psi}{\partial t}$. For our case, if the external metric $f_{\mu\nu} $ depends explicitly on time , we should include terms coming  from explicit time dependence of $f_{\mu\nu} $ present in the expression of $\chi$. As so many references, we do not consider such cases here.}
\begin{equation}
\lbrace \chi(x),H_{0} \rbrace _{*}+\int d^{3}y N(y)\lbrace \chi(x),\chi(y) \rbrace_{*}=0, \label{dbb} 
\end{equation}
in which the symbol $\lbrace \ ,\ \rbrace_{*}$  means the Dirac bracket defined as
\begin{equation}
\lbrace f,g \rbrace_{*}=\lbrace f,g \rbrace+\lbrace f,\xi_{a} \rbrace C^{ab}\lbrace \xi_{b},g \rbrace,\label{dbbc} 
\end{equation}
where $C^{ab}$ is the inverse of the matrix of Poisson brackets of second class constraints $\xi_{a}\equiv (p_{i},\vartheta_{i})$.
In Eq. (\ref{dbbc}) summations on the repeated indices include integration on the three dimensional space variables.  Since the constraint $\chi$ contains spatial derivatives of the variables, the second term of Eq. (\ref{dbb})    does not vanish trivially. Using the antisymmetry property of  $C^{ab}$ it is easily seen that 

\begin{equation}
\lbrace \chi(x),\chi(y) \rbrace_{*}=\lbrace \chi(x),\chi(y) \rbrace\approx0. \label{equal}
\end{equation}
Hence Eq. (\ref{dbb}) can be written as
\begin{equation}
\varUpsilon+\int d^{3}y N(y)\lbrace \chi(x),\chi(y) \rbrace=0, \label{opsi}
\end{equation}
where $\varUpsilon\equiv\lbrace \chi(x),H_{0} \rbrace_{*}$. Suppose at this stage we are given a generic interaction term which can be linearised with respect to the lapse function as (see Eq. (\ref{HRL}))
\begin{equation}
V_{int}=2m^{2}\sqrt{\gamma}(MK+NZ). 
\end{equation}
For our case (i.e. action (\ref{HRa})) the functions $K$ and $Z$ are given in Eq. (\ref{INT1}), Eq. (\ref{INT2}). It can be shown directly 
that consistency condition lead to the following equation 
\begin{eqnarray}
\varUpsilon-2\tau^{i}\partial_{i}N -N\partial_{i}\tau^{i}\approx0,
\end{eqnarray}
where 
\begin{eqnarray}
\tau^{i}= (R_{0}+R_{m}D_{j}^{m}n^{j})D_{l}^{i}n^{l}+R^{i}
+2\gamma_{lj}D_{k}^{j}n^{k} \left( R_{m} \frac{(\partial D_{r}^{m}n^{r}) }{\partial \gamma_{il}} +2m^{2}\frac{\partial (\sqrt{\gamma}Z)}{\partial \gamma_{il}} \right). \label{pi1}
\end{eqnarray}

If $\tau^{i}$ are non zero on the constraint surface, the left hand side of Eq. (\ref{opsi}) reads as a constraint which includes the lapse function $N$ linearly. Consistency of such a constraint with the total Hamiltonian $\mathcal{H}_{T}=\mathcal{H}_{c}+uP$ leads to determination of the remaining Lagrange multiplier $u$, and the consistency process would stop at this point. If this is the case, we will not have enough constraints to omit the ghost degree of freedom. Hence, in order to have a ghost free model we need another condition on the $Z$ part of the interaction term as $\tau^{i}\approx0$. 
This criterion is a simple explanation of the criterion given in Ref. \cite{comeli1} (see equation (4.11) therein), for the case where the interaction term is linearized with respect to lapse function.
For our current case where $Z$ is given by equation (\ref{INT2}) it can be shown directly that 
\begin{equation}
\tau^{i}= \chi(x)D_{j}^{i}n^{j}(x), \label{equal2}
\end{equation}
which vanishes weakly due to $\chi \approx 0$. 
Turning back to Eq. (\ref{opsi}) the Dirac bracket $ \lbrace \chi, H_{0}\rbrace_{*} $ can be achieved by the usual Poisson bracket in which (in addition to explicit dependence) the implicit dependence of the expressions on the canonical variables through dependence of $ n^{i} $ on $ \gamma_{ij} $ and $ \pi^{ij} $, should also be taken into account, as follows 
\begin{eqnarray}&&
\lbrace \chi, H_{0}\rbrace_{*}=(\frac{\delta \chi }{\delta \gamma_{mn}}+\frac{\delta \chi }{\delta n^{i}}\frac{\delta n^{i} }{\delta \gamma_{mn}})(\frac{\delta H_{0} }{\delta \pi^{mn}}+\frac{\delta H_{0} }{\delta n^{i}}\frac{\delta n^{i} }{\delta \pi^{mn}})\nonumber \\&&\hspace{17mm}-(\frac{\delta \chi }{\delta \pi^{mn}}+\frac{\delta \chi }{\delta n^{i}}\frac{\delta n^{i} }{\delta \pi^{mn}})(\frac{\delta H_{0} }{\delta \gamma_{mn}}+\frac{\delta H_{0} }{\delta n^{i}}\frac{\delta n^{i} }{\delta \gamma_{mn}}) \label{a32},
\end{eqnarray}
However, since $\frac{\delta \chi }{\delta n^{i}}=\frac{\delta (D^{k}_{j}n^{j})}{\delta n^{i}}\vartheta_{k}\approx 0$ and $\frac{\delta H_{0} }{\delta n^{i}}=\vartheta_{i} \approx 0$, the Dirac bracket (\ref{a32}) reduces weakly to ordinary Poisson bracket $\lbrace \chi, H_{0}\rbrace$. The final result for the third level constraint reads \cite{mmg2}
\begin{eqnarray}&&
\varUpsilon\hspace{3mm}=m^{2}M(\gamma_{mn}\pi-2\pi_{mn})(\frac{2}{\sqrt{\gamma}}\frac{\partial (K\sqrt{\gamma})}{\partial \gamma_{mn}})\nonumber \\&& \hspace{10mm}+2m^{2}M\sqrt{\gamma}\triangledown_{m}(\frac{2}{\sqrt{\gamma}}\frac{\partial K\sqrt{\gamma}}{\partial \gamma_{mn}})\gamma_{ni}D^{i}_{k}n^{k}
+\left(R_{j}D^{i}_{k}n^{k}\right)\triangledown_{i}(Mn^{j}+M^{j})\nonumber \\&&\hspace{10mm}
-2m^{2}\sqrt{\gamma}\gamma_{jk}( \gamma^{ki^{\prime}} ( \beta_{1}(x)^{-1/2}f_{i^{\prime}k^{\prime}} (D^{-1})^{k^{\prime}}_{j^{\prime}}+\beta_{2}(f_{i^{\prime}k^{\prime}}(D^{-1})^{k^{\prime}}_{j^{\prime}}D-f_{i^{\prime}j^{\prime}}) \hspace{15mm}\nonumber \\&& \hspace{10mm}
+\beta_{3}\sqrt{x}( f_{i^{\prime}k^{\prime}}D^{k^{\prime}}_{j^{\prime}}-f_{i^{\prime}j^{\prime}}D
+\frac{1}{2} f_{i^{\prime}k^{\prime}}(D^{-1})^{k^{\prime}}_{j^{\prime}}(D^{2}-D^{l^{\prime}}_{h^{\prime}}D^{h^{\prime}}_{l^{\prime}}))    ) \gamma^{j^{\prime} i})\triangledown_{i}(Mn^{j}+M^{j})\nonumber \\&&\hspace{10mm}+\sqrt{\gamma}(Mn^{i}+M^{i})[\triangledown_{i}(\frac{R^{0}}{\sqrt{\gamma}})+
\triangledown_{i}(\frac{R_{j}}{\sqrt{\gamma}})D^{j}_{k}n^{k}]. \label{op}
\end{eqnarray}
Consistency condition of  $\varUpsilon(\pi_{ij},\gamma^{ij})$ then gives
\begin{equation}
\varOmega\equiv \lbrace \varUpsilon, H_{0}\rbrace_{*}+N \lbrace \varUpsilon,\chi \rbrace_{*}  \approx 0. \label{omega}
\end{equation}
Since $\varOmega$ depends linearly on $N$,  consistency of $\varOmega$ determines the remaining Lagrange multiplier $u$ in the total Hamiltonian. In this way the canonical investigation of the system comes to its end with 10 second class constraints.

 \end{document}